\newif\iftwocol

\documentclass[final,5p,times,twocolumn,authoryear]{elsarticle}
\twocoltrue

\usepackage{amssymb}

\usepackage{threeparttable}
\usepackage{floatpag}

\newcommand{\um}{\ensuremath{\,\mu{\rm m}}}
\newcommand{\odeg}{\ensuremath{^\circ }}
\newcommand{\methane}{\ensuremath{{{\rm CH}_4}}}

\def\aj{{AJ} }                   
\def\apjl{{ApJ} }                
\def\icarus{{Icarus} }           
\def\nat{{Nature} }              
\def\grl{{Geophys.~Res.~Lett.} } 
\def\jgr{{J.~Geophys.~Res.} }    
\def\jqsrt{{J.~Quant.~Spec.~Radiat.~Transf.} } 
\def\planss{{Planet.~Space~Sci.} }   

\journal{Icarus}
\begin{document}
\begin{frontmatter}

\title{Meridional variation in tropospheric methane on Titan \\ observed with AO 
spectroscopy at Keck and VLT}

\author[label1]{M\'at\'e~\'Ad\'amkovics}
\ead{mate@berkeley.edu}
\ead[url]{http://astro.berkeley.edu/~madamkov}

\author[label2a,label2b]{Jonathan~L.~Mitchell}
\author[label3]{Alexander~G.~Hayes}
\author[label4]{Patricio~M.~Rojo}
\author[label3]{Paul~Corlies}
\author[label5]{Jason~W.~Barnes}
\author[label5b]{Valentin~D.~Ivanov}
\author[label6]{Robert~H.~Brown} 
\author[label7]{Kevin~H.~Baines}
\author[label8]{Bonnie~J.~Buratti} 
\author[label9]{Roger~N.~Clark} 
\author[label10]{Philip~D.~Nicholson} 
\author[label8]{Christophe~Sotin}

\address[label1]{Astronomy Department, University of California, Berkeley, CA 94720-3411, USA} 
\address[label2a]{Department of Earth \& Space Sciences, \\
University of California Los Angeles, Los Angeles, CA 90095, USA}
\address[label2b]{Department of Atmospheric \& Oceanic Sciences, \\
University of California Los Angeles, Los Angeles, CA 90095, USA} 
\address[label3]{Center for Radiophysics and Space Research, \\ 
				 Cornell University, Ithaca, NY 14853, USA}
\address[label4]{Universidad de Chile, Camino El Observatorio 1515, 
			     Las Condes, Casilla 36-D Santiago, Chile}
\address[label5]{Department of Physics, University of Idaho, Moscow, ID 83844-0903, USA}
\address[label5b]{European Southern Observatory, Ave. Alonso de Cordova 3107,
                  Casilla 19001, Santiago, Chile}
\address[label6]{Lunar and Planetary Laboratory, University of Arizona, Tucson, AZ 85721, USA.}
\address[label7]{Space Science and Engineering Center, \\ 
				 University of Wisconsin, Madison, WI 53706, USA.}
\address[label8]{Jet Propulsion Laboratory, Caltech, Pasadena, CA, 91109 USA}
\address[label9]{United States Geological Survey, Denver, CO, 80225 USA}
\address[label10]{Department of Astronomy, Cornell University, Ithaca, NY 14853 USA}

\begin{abstract}

The spatial distribution of the tropospheric methane on Titan was measured using near-infrared
spectroscopy. Ground-based observations at 1.5\um\ (H-band) were performed during the same night
using instruments with adaptive optics at both the W.~M.~Keck Observatory and at the Paranal
Observatory on 17~July~2014~UT. The integral field observations with SINFONI on the VLT covered the
entire H-band at moderate resolving power, $R=\lambda/\Delta\lambda\approx1,500$, while the Keck
observations were performed with NIRSPAO near 1.5525\um\ at higher resolution, $R\approx25,000$. The
moderate resolution observations are used for flux calibration and for the determination of model
parameters that can be degenerate in the interpretation of high resolution spectra. Line-by-line
calculations of CH$_4$ and CH$_3$D correlated $k$ distributions from the HITRAN 2012 database were
used, which incorporate revised line assignments near 1.5\um. We fit the surface albedo and aerosol
distributions in the VLT SINFONI observations that cover the entire H-band window and used
these quantities to constrain the models of the high-resolution Keck NIRSPAO spectra when retrieving
the methane abundances. {\it Cassini} VIMS images of the polar regions, acquired on 20 July 2014 UT,
are used to validate the assumption that the opacity of tropospheric aerosol is relatively uniform
below 10\,km. We retrieved methane abundances at latitudes between 42\odeg S and 80\odeg N. 
The tropospheric methane in the Southern mid-latitudes was enhanced by a factor of $\sim$10--40\% over
the nominal profile that was measured using the GCMS on {\it Huygens}. The Northern hemisphere had
$\sim$90\% of the nominal methane abundance up to polar latitudes (80$^{\circ}$N). These measurements
suggest that a source of saturated polar air is equilibrating with dryer conditions at lower
latitudes.

\end{abstract}


\tnotetext[t1]{Accepted for publication on May 22, 2015}

\end{frontmatter}


\section{Introduction}\label{s:Intro}

Methane (\methane) is the most abundant condensible species on Titan, dominates the energy
transport through the atmosphere \citep{Mitchell2012}, and is part of a complex hydrological cycle
\citep{Atreya2006,Roe2012}. Clouds of methane can indicate regions of convection
\citep[e.g.,][]{Griffith2005}, polar subsidence \citep{Anderson2014}, or evaporation from lakes
\citep[e.g.,][]{Brown2009b, Turtle2009}, while the formation of large scale methane cloud systems
are diagnostic of atmospheric dynamics via their morphology \citep{Mitchell2011} and how they evolve
with time \citep{Adamkovics2010, Turtle2011a}. The amount of methane near the surface is an
important factor in triggering convective cloud formation \citep{Barth2007} and in determining the
strength of storms \citep{Hueso2006}. Precipitation can  return methane to the surface
\citep{Turtle2009,Turtle2011a} where fluid transport has some role in closing the hydrological
cycle. Seasonal variations in the general circulation \citep{Mitchell2009b} as well as predictions
of the locations and frequency of clouds \citep{Schneider2012} depend on the distribution of methane
near the surface, both in  the regolith and the lower atmosphere.

Lakes and seas of liquid hydrocarbons \citep{Stofan2007} provide both sinks and sources of methane
on the surface.  The north pole contains by far the greatest extent of open liquids on Titan
\citep{Hayes2008, Lorenz2008,Sotin2012,Lorenz2014}.  The largest sea, Kraken Mare, extends down to
$55^\circ$\,N at its southernmost point.  A few low-latitude lake candidates have been suggested,
one near the equator \citep{Griffith2012a}, and one at $40^\circ$\,S latitude \citep{Vixie2014}.
The sole large lake in Titan's south polar region is Ontario Lacus \citep{Turtle2009}, although
there are several basins that have been identified as potential "paleo-seas" that encompass a
similar areal fraction to the northern seas \citep{Hayes2011}.

Liquids presumably concentrate at the poles because they are the coldest points on the surface and
therefore cold-traps for volatiles.  However, the polar clustering might also be related to higher
precipitation at the poles \citep{Rodriguez2009,Brown2010,Rodriguez2011} relative to the dune-filled
equatorial desert \citep{Lorenz2006a,Radebaugh2008,LeGall2011,Rodriquez2014}, which may be caused by
circulation \citep{Rannou2006,Friedson2009}.  The lower elevations of the poles relative to
equatorial regions may also play a role \citep{Iess2010,Lorenz2013}. The reason for the pronounced
North-South asymmetry in lake coverage is unknown. \citet{Aharonson2009} cited Milankovic-like
cycles in Titan's orbital parameters as a possible explanation, which is supported by simulations of
Titan's paleoclimate \citep{Lora2014}.

The physical properties of lakes are complicated by the fact that they are likely mixtures of
hydrocarbons. Though methane composes Titan's raindrops, the seas may build up significant fractions
of less-volatile ethane. Spectroscopic observations of Ontario Lacus suggest the presence of ethane
\citep{Brown2008}, although the abundance is not constrained by these measurements.
Recent observations of Ligeia Mare, conducted by the Cassini RADAR instrument, have demonstrated
that it is primarily composed of methane \citep{Mastrogiuseppe2014}. Evaporation rates from a lake
that is mostly methane will be much greater than from a lake that is mostly ethane
\citep{Mitri2007,Tokano2009}. \citet{Lorenz2014a} points out that the ratio of methane to ethane may
vary across lakes due to the concentration of solutes by heterogeneous evaporation and dilution by
heterogeneous rainfall.

While shoreline recession at Ontario Lacus was reported over the timescale of the {\it Cassini}
mission \citep{Turtle2011b,Hayes2009,Hayes2011}, the shoreline detection algorithms have been
disputed \citep{Cornet2012}, leaving the contemporary evaporation rate over lakes uncertain. {\it
Cassini} has observed albedo variations with both the Imaging Science Subsystem (ISS) and
RADAR that depict smaller southern lakes disappearing between adjacent observations
\citep{Turtle2009,Hayes2011}, which was attributed to either infiltration or evaporation, although
the rates could not be quantified. Changes in lake and sea volumes over geologic timescales have
also likely occurred,  as evidenced by the geomorphology of empty lakebeds in some polar terrains
and the presence of drowned river valleys in the northern seas, which indicate that the liquid level
is rising faster than fluvial sediment is being deposited
\citep{Hayes2008}. Some of the lakebeds show a  bright reflection near 5\um, which is interpreted as
a compositional signature that is attributed to the formation of organic evaporite
\citep{Barnes2011}.  The largest outcrop of evaporites are in the tropics, implying that these areas
may have been seas during the geological past, perhaps under a different climatic regime
\citep{Moore2010,MacKenzie2014}. 

The evaporation of methane from surface lakes may have an observational impact on the atmosphere.
\citet{Tokano2014} recently revisited the {\it Cassini} radio occultation data (recorded from
2005--2009) and points out that retrievals assuming a uniform tropospheric methane distribution lead
to surface pressure distributions that are inconsistent with the predictions of circulation models.
Instead, \citet{Tokano2014} argues for a substantially higher methane abundance in the Summer
hemisphere. \citet{Penteado2010a} searched for spatial variation in the methane abundance with high
resolution Keck observations. The unsaturated lines of the resolved 3$\nu_2$ band of CH$_3$D are
sensitive to possible changes in the tropospheric methane abundance. Their measurements from
December 2006 indicated that methane below 10\,km altitude is constant to within 20\% in the
tropical atmosphere, sampled between the range of 32\odeg S--18\odeg N. High resolution analysis
with new methane lines lists \citep{deBergh2012}  illustrates that significant improvements can be
made in spectral fitting with recent laboratory data.

Here we present ground-based observations of the methane distribution on Titan using a methodology
that improve upon the observing protocol and integration times of \citet{Penteado2010a}, and which
are supported by both integral field observations from the same night, as well as a {\it Cassini}
flyby from four days later. Our radiative transfer models include revised methane line lists from
the most recent HITRAN database. The observations, data reduction, and calibrations are described in
Section~\ref{s:Observations}, while the radiative transfer model is detailed in Section~\ref{s:RT}.
Results are presented in Section~\ref{s:Results} and discussed in Section~\ref{s:Discussion}.

\begin{figure}[t]\begin{center}
\iftwocol
	\includegraphics[width=3.5in]{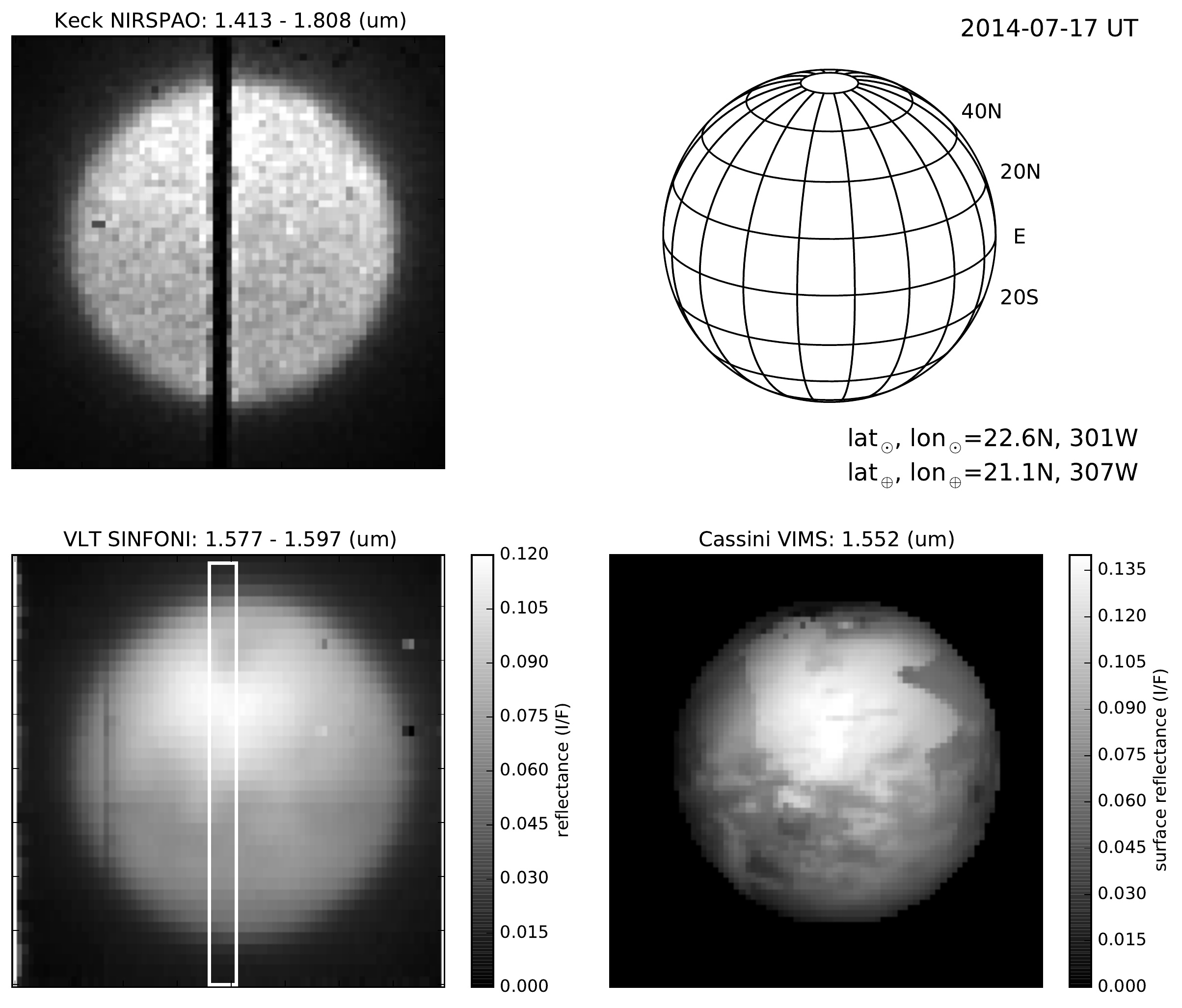} 
\else
	\includegraphics[width=5.25in]{f1}  
\fi
\caption{\label{f:view}Viewing geometry and surface albedo of Titan during observations.
An H-band image from the NIRSPAO slit viewing camera (top left) during the spectrometer exposures
identifies the spatial location of the spectra. The wide bandpass results in sensitivity to the
atmospheric haze. An image from the SINFONI data cube (bottom left) that is sensitive to surface
albedo variations is shown with a white box that indicates the region of pixels that corresponds to
the NIRSPAO slit. This is the region considered when referring to pixels along the SINFONI ``slit".
An orthographic reprojection of a 1.552\um\ {\it Cassini} Visual and Infrared Mapping Spectrometer
(VIMS) map, sampled at the SINFONI plate scale and artificially limb-darkened, illustrates the
surface reflectance in the absence of the atmospheric contribution and confirms the photometry and
flux calibration of the SINFONI data (bottom right). The viewing geometry of the observations is
show in the schematic (top right) with a grid spacing of 20 degrees. }
\end{center} \end{figure}

\section{Observations}\label{s:Observations}

Observations were performed on 17~July~2014~UT at both the Paranal Observatory and the W.~M. Keck
Observatory. Instrumentation with complementary observing modes, resolutions, and bandpasses
provided flux calibration and characterization of the physical properties of the atmosphere and
surface. Figure~\ref{f:view} illustrates the viewing geometry of the observations that are described
below.

\subsection{VLT Observations}

The Spectrograph for INtegral Field Observations in the Near Infrared (SINFONI) on the Very Large
Telescope (VLT) at Paranal Observatory was used as part of a campaign to monitor clouds on Titan.
The spectrometer is fed by an adaptive optics module and uses two sets of stacked mirrors to
optically divide and rearrange the field of view (FOV) into a single synthetic long slit that is
spectrally dispersed by a grating onto the  detector. We used the 0.8"$\times$0.8" FOV,
corresponding to a spatial pixel scale of 0.0125"$\times$0.0250",  with the grating that covers 1.45
-- 2.45\um\ at a spectral resolution of $\Delta \lambda \approx 1$\,nm,  corresponding to a
resolving power, $R  =\lambda / \Delta \lambda \approx 1500$ \citep{Eisenhauer2003c}. Four
overlapping exposures with 2 $\times$ 15\,sec coadds, offset by $\pm$0.1" from the disk center in
both the X and Y directions of the FOV, are mosaicked to cover the entire disk.

Observations were reduced using version 2.5.2 of the SINFONI pipeline. The standard processing of
the raw exposures includes correction of bad pixels, flat fielding, and correction of geometric
distortions. The pipeline performs wavelength calibration and then reconstructs the 32 slices of
spectra into a data cube. Sky frames are used to correct for sky emission.

The B3V type star Hip~74680 (HD~134485) was observed at an airmass of 1.019 and is used for both
telluric correction and flux calibration. A high resolution H-band telluric template from ESO is
convolved to the instrument resolution and scaled  to fit telluric absorption features observed in
the calibration star near 1.47 and 1.58\um; the normalized  telluric template was scaled by a factor
of 0.7 (and offset by 0.3), and then used to correct the target spectra.

Photometric calibration was performed by integrating the calibration star spectrum over the 2MASS
filter curves in H-band and comparing with the known apparent magnitude. The relative spectral
response $R_{\rm H}(\lambda )$, was used to determine the observed isoflux,
\begin{equation} 
I_{\rm H} = \frac{\int R_{\rm H}(\lambda ) I_{\rm obs}(\lambda) 
			d\lambda}{\int R_{\rm H}(\lambda ) d\lambda},
\end{equation}  
where $I_{\rm obs}(\lambda)$ is the observed spectrum. The 2MASS ``zero-magnitude'' reference flux
in H-band from \citet{Cohen2003} is $$ I_{\rm ref} = 1.133 \times 10^{-13} \pm 2.212\times 10^{-15}
\; {\rm W \, cm^{-2}\, \mu m^{-1} }. $$ The spectral bandpass of SINFONI covers all of H-band,
including the entire 2MASS H-band filter range, which facilitates the calibration of the NIRSPAO
observations that cover a narrow band-pass at higher resolving power.

The apparent H-band magnitude of the calibrator is $m_{\rm H}=8.322$ from 2MASS \citep{Cutri2003},
and the correction factor for converting the observed photon count rate (DN/s) to flux is given by
\begin{equation} 
c_{\rm H} = \frac{I_{\rm ref}}{I_{\rm H}}10^{-m_{\rm H}/2.5}.
\end{equation}  
The photometrically calibrated spectra, $ I(\lambda) = I_{\rm obs}(\lambda)c_{\rm H}  $, were then 
converted to units of reflectivity,
\begin{equation}
\frac{I}{F} = \frac{r^2}{\Omega}\frac{I(\lambda)}{F_\odot (\lambda)},
\end{equation}
where $r$ is the heliocentric distance to Titan, $\Omega$ is the solid angle (in steradians) of each
spatial pixel, and $\pi{F_\odot}(\lambda)$ is the solar spectrum at 1\,AU, for which the 2000 ASTM
Standard Extraterrestrial Spectrum Reference\footnote{http://rredc.nrel.gov/solar/spectra/am0/} was
used.

\begin{table*}
	\centering
	\begin{threeparttable}
	\caption{Log of observations on 2014-07-17}
	\label{t:obslog}
	\begin{tabular}{lccccc}
\hline \hline
Target & Instrument & Time (UTC) &  Exposure (s) & Airmass & Lon.\tnote{a} \\
\hline  Titan & NIRSPAO & 05:49:30 & 300 & 1.212 &  290.6 \\
Titan & NIRSPAO & 05:55:33 & 300 & 1.214 &  290.7 \\
Titan & NIRSPAO & 06:00:47 & 300 & 1.216 &  290.7 \\
Titan & NIRSPAO & 06:07:56 & 300 & 1.220 &  290.9 \\
Titan & NIRSPAO & 06:16:28 & 300 & 1.227 &  291.0 \\
Titan & NIRSPAO & 06:23:20 & 300 & 1.234 &  291.1 \\
Titan & NIRSPAO & 06:28:34 & 300 & 1.240 &  291.2 \\
Titan & NIRSPAO & 06:34:13 & 300 & 1.248 &  291.3 \\
Titan & NIRSPAO & 06:40:19 & 300 & 1.257 &  291.4 \\
Hip 77516 & NIRSPAO & 08:30:03 & 20 & 1.254 & -- \\ 
Titan & SINFONI & 23:39:31 &  30   &  1.018 &  307.3  \\
Titan & SINFONI & 23:40:32 &  30   &  1.018 &  307.4  \\
Titan & SINFONI & 23:41:31 &  30   &  1.018 &  307.4  \\
Titan & SINFONI & 23:42:29 &  30   &  1.018 &  307.4  \\
Hip 74680 & SINFONI & 23:53:21 & 4 & 1.019  &   --    \\
\hline\\
	\end{tabular}
	\begin{tablenotes}
	\footnotesize
	\item[a] Sub-observer longitude from JPL Horizons ephemerides.
	\end{tablenotes}
	\end{threeparttable}
\end{table*}

\subsection{Keck Observations}

The Near-InfraRed SPECctrometer (NIRSPEC) on the Keck~II telescope at W.~M.~Keck Observatory 
\citep{McLean1998} was used with the adaptive optics (AO) system, NIRSPEC+AO (NIRSPAO),
for high resolution spectroscopy of Titan with a pixel scale of 0.018"/pixel along the slit. The
instrument was setup in the cross-dispersed echelle mode in H-band with a 0.041"$\times$2.26" slit,
giving a resolving power of $R\approx25,000$. The echelle and cross-disperser angles were set to
62.8 and 36.5, respectively, nearly continuously covering the range 1.481 -- 1.701\um\ in 7 echelle
orders. The edges of neighboring orders can overlap with the 2.26" slit in this setting. For this
work we focused on Order 49, sampling the 1.541 -- 1.563\um\ region, and where there was significant
overlap at the edges of the order.

The standard ABBA dither (nod) pattern for the instrument moves the target 25\% of the length of the
slit (0.57") from center, which would position the limb of the $\sim$0.8" disk of Titan near the
overlapping region of echelle order. This was avoided by using a smaller dither step along the slit.
Sky exposures taken completely off target were recorded for sky subtraction. The integration time
for each spectrum was 300\,s. Two ABBA sets were record with the slit aligned North-South near the
central meridian, and one exposure was taken with Titan in the center of the slit, for a total of
2,700\,s integration on target, followed by a 300\,s sky exposure. The spectral type A0V calibrator
star was HD~141513 (Hip 77516). The logs for both SINFONI and NIRSPAO observations are presented in
Table~\ref{t:obslog}.

The NIRSPAO data are reduced using standard procedures for bad pixel correction, flat fielding, and
cosmic ray rejection. The spatial and spectral rectification routines from the REDSPEC package are
used before shifting and stacking exposures. A bare sky exposure is used for first order sky
subtraction. A second order correction was performed by (1) offsetting each exposure to the median
of the stacked set near the edge of the order, and then (2) scaling each exposure to the median
value at the center of the order. The stack of images for individual exposures was collapsed using
the mean of each pixel.

\begin{figure}[]\begin{center}
\iftwocol
	\includegraphics[width=3.5in]{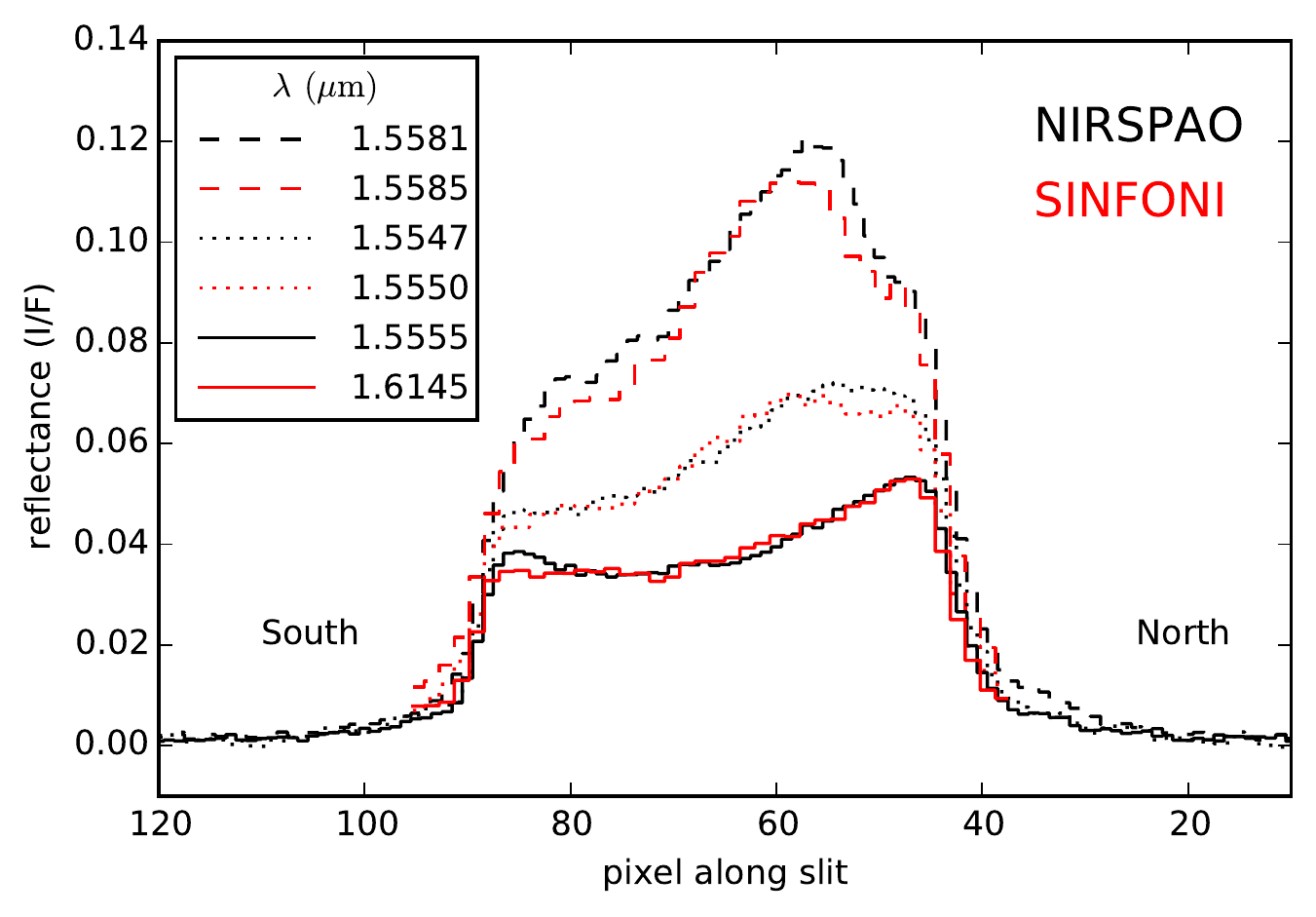}  
\else
	\includegraphics[width=4.5in]{f2}  
\fi
\caption{\label{f:fluxcal} Reflectance profiles along the NIRSPAO slit (black) and 
corresponding SINFONI pixels (red) at wavelengths that are sensitive to the surface (dashed), 
and atmosphere (dotted and solid). The profiles that are sensitive to the
surface are limb-darkened whereas the profiles sensitive to the atmosphere exhibit 
limb-brightening. The spatial region of pixels that is used from the SINFONI cube is 
illustrated by the white box in the bottom left panel of Figure~\ref{f:view}. 
The SINFONI pixel scale has been linearly mapped to the NIRSPAO pixel scale.
The difference between reflectivity profiles of the surface (red and black dashed curves) is 
consistent with the surface albedo changing due to the rotation between the time of the 
two observations.
}
\end{center} \end{figure}

Flux calibration of an extended object in a slit spectrometer can be challenging because the slit
losses due to the point spread function (PSF) extending over the edge of the slit need to be
determined. For point sources, it can often be assumed that the unknown slit losses for the
calibrator and the target are the same. However, this is not the case when comparing a point source
calibrator and an extended target. After an initial telluric and flux calibration was performed
using Hip~77516 (HD~141513), the NIRSPAO data were then scaled to match the calibrated SINFONI
observations. Two wavelengths were considered that are sensitive to the surface and lower
atmosphere, Figure~\ref{f:fluxcal}.

Unlike the narrow NIRSPAO channels, which can probe regions of high methane opacities near 1.55\um\
and therefore the atmosphere, the SINFONI channels near 1.55\um\ are broad, cover regions of
predominantly low methane opacity, and are therefore more sensitive to the surface. Due to the
difference in bandpass, the reflectance along the NIRSPAO slit at 1.5555\um\ does not match the
corresponding SINFONI channel and is compared to the 1.6145\um\ SINFONI channel in
Figure~\ref{f:fluxcal}. The 1.55\um\ spectral region is generally thought to be a surface-sensitive
on Titan, however, when sampled at high spectral resolution there are wavelengths in this region
with large gas opacity that probe only the atmosphere.

\subsection{{\it Cassini} VIMS Observations}

Spectral mapping cubes were obtained by {\it Cassini} VIMS \citep{Brown2004} during both the ingress
and egress of the T103 flyby on 2014~Jul~20~UT, and provide views of both polar regions. We reduce
the VIMS IR channels from two cubes (datasets 1784502376\_1 and 1784584782\_1) using the standard
pipeline for calibration and determination of the viewing geometry. Images from the 1.573\um\ and
1.625\um\ channels are used to evaluate meridional variation in near-surface hazes toward both
poles. Since the tropospheric haze opacity near the surface can be degenerate with the methane
abundance there (described below) we use the VIMS observations to differentiate between models.

\section{Radiative Transfer}\label{s:RT}

We implement a model of the atmosphere using the {\em in situ} measurements made with instruments on
the {\it Huygens} probe, which provide  structure, chemical composition, and aerosol scattering in
our reference model. Methane line assignments from the HITRAN 2012 database \citep{Rothman2013} are
used to determine gas opacities via line-by-line methods and we solve the radiative transfer using
the discrete ordinates  method \citep{Stamnes1988}.

\subsection{Structure and Composition}

We use a model with 20 layers, which have  boundaries (levels) that are evenly spaced in pressure
above and below 300\,mbar (hPa), covering the 2.75 -- 1466.45\,mbar range sampled by the {\it
Huygens} Atmospheric Structure Instrument (HASI) on {\it Cassini} \citep{Fulchignoni2005}. This
corresponds to an altitude range from the surface up to 147\,km. There are 10 layers sampled at
$\sim$30\,mbar intervals through the tropopause and stratosphere and 10 layers sampling the
troposphere at $\sim$115\,mbar steps. Fewer layers can improve the computational speed, but yield
inconsistent calculations, whereas increasing the number of layers above 20 has no significant
benefit. Altitude sensitivity for cloud retrievals may be improved with additional layers, but this
topic is beyond the scope of this work. After the levels in the model are determined, the pressures,
temperatures and densities for each layer are interpolated and the total column densities are
determined using the mole fractions reported by the {\it Huygens} Gas Chromatograph Mass
Spectrometer (GCMS) \citep{Niemann2010}. The atmospheric structure and composition are tabulated in
Table~\ref{t:struct}. \begin{table*}[ht]
	\centering
	\begin{threeparttable}
	\caption{Atmospheric Structure\tnote{a} and Composition\tnote{b}}
	\label{t:struct}
	\begin{tabular}{*{8}{c}} 
\hline\hline
 Layer & Pressure & Altitude & Temperature & Density & \multicolumn{2}{c}{Total Column} & CH$_4$ Column \\ 
   & (mbar) & (km) & (K) & (cm$^{{-3}}$) & (cm$^{{-2}}$) & (km amg) & (km amg) \\ \hline 
  1&  9.5   & 105.8  & 121.1  &5.75e+17 &5.14e+24 &  1.91  & 0.0286  \\ 
  2&  44.9  &  59.1  &  79.1  &4.13e+18 &4.89e+24 &  1.82  & 0.0272  \\ 
  3&  75.6  &  50.4  &  70.9  &7.73e+18 &4.87e+24 &  1.81  & 0.0272  \\ 
  4& 105.7  &  45.0  &  70.5  &1.09e+19 &4.87e+24 &  1.81  & 0.0272  \\ 
  5& 135.7  &  41.1  &  70.5  &1.40e+19 &4.86e+24 &  1.81  & 0.0275  \\ 
  6& 165.6  &  37.9  &  70.7  &1.71e+19 &4.85e+24 &  1.81  & 0.0282  \\ 
  7& 195.4  &  35.3  &  71.0  &2.00e+19 &4.85e+24 &  1.81  & 0.0293  \\ 
  8& 225.2  &  33.0  &  71.5  &2.29e+19 &4.83e+24 &  1.80  & 0.0302  \\ 
  9& 255.0  &  31.1  &  72.0  &2.58e+19 &4.85e+24 &  1.80  & 0.0313  \\ 
 10& 284.7  &  29.3  &  72.6  &2.86e+19 &4.84e+24 &  1.80  & 0.0323  \\ 
 11& 353.5  &  25.7  &  74.1  &3.49e+19 &1.90e+25 &  7.07  & 0.1398  \\ 
 12& 471.4  &  20.9  &  76.6  &4.51e+19 &1.89e+25 &  7.05  & 0.1665  \\ 
 13& 588.7  &  17.1  &  78.9  &5.47e+19 &1.88e+25 &  7.00  & 0.1942  \\ 
 14& 705.9  &  13.9  &  81.1  &6.38e+19 &1.89e+25 &  7.02  & 0.2350  \\ 
 15& 822.8  &  11.1  &  83.1  &7.25e+19 &1.88e+25 &  7.00  & 0.2831  \\ 
 16& 939.7  &  8.7   &  84.9  &8.10e+19 &1.87e+25 &  6.96  & 0.3363  \\ 
 17& 1056.6 &  6.5   &  86.8  &8.92e+19 &1.87e+25 &  6.98  & 0.3754  \\ 
 18& 1173.4 &  4.5   &  88.7  &9.72e+19 &1.88e+25 &  7.01  & 0.4002  \\ 
 19& 1290.2 &  2.6   &  90.6  &1.05e+20 &1.88e+25 &  7.01  & 0.4027  \\ 
 20& 1406.9 &  0.8   &  92.5  &1.12e+20 &1.89e+25 &  7.04  & 0.3514  \\ 
\hline
	\end{tabular}
	\begin{tablenotes}
	\footnotesize
	\item[a] Pressures, temperatures, and densities are interpolated from the 
	HASI on Cassini \citep{Fulchignoni2005}. 
	\item[b] \methane\ mole fractions are from the Huygens GCMS \citep{Niemann2005,Niemann2010}.
	\end{tablenotes}
	\end{threeparttable}
\end{table*}

\subsection{Aerosol Model}

The aerosol scattering phase functions and opacities were measured {\em in situ} by the the Descent
Imager-Spectral Radiometer (DISR) on the {\it Huygens} probe \citep{Tomasko2008c}. We fit 32nd order
Legendre polynomials to the phase functions tabulated at 1.29\um, 1.58\um, and 2.00\um, using a
Levenberg-Marquardt (LM) optimization, and linearly interpolate coefficients for intermediate
wavelengths. The vertical opacity structure from the model of \citep{Tomasko2008c} is used, with a
haze single scattering albedo, $\omega_H=0.96$.

\subsection{CH$_4$ and CH$_3$D opacities}

Spectra resolving the natural line shape of CH$_4$ and CH$_3$D are calculated using line-by-line
methods. These spectra are used to calculate correlated-$k$ coefficients at the resolution and
dispersion plate scale of both the NIRSPAO and SINFONI instruments. Correlated-$k$ values
\citep{Lacis1991} for \methane\ and CH$_3$D are calculated for 390 combinations of temperature and
pressure that have been used in the literature, e.g., by \citet{Sromovsky2012} and
\citet{Irwin2014}. \methane\ and CH$_3$D lines are from the HITRAN 2012 database
\citep{Rothman2013}, which include the WKMC-80K methane line data of \citet{Campargue2012} that are
relevant to this spectral region.

We calculate the monochromatic opacity at frequency, $\nu$, pressure $P$, and temperature, $T$, by
summing over lines, $\ell$, as described by
\citet{Sromovsky2012},
\iftwocol \footnotesize \fi
\begin{equation}
 k(\nu; P, T) =  \sum_\ell  S_\ell(T_0) \exp \left[ \frac{hcE_\ell}{k_B}  \left( \frac{1}{T_0} 
 - \frac{1}{T} \right)\right] \frac{Q(T_0)}{Q(T)} f_\ell(\nu-\nu_\ell; P,T), 
\end{equation}
\iftwocol \normalsize \fi
correcting the typo of the reversed $T_0$ and $T$ in the square brackets of their Equation 1.
$S_\ell(T_0)$ is the line strength at reference temperature $T_0$, $E_\ell$ is the lower state
energy of the line, and the partition function ratio $Q(T_0)/Q(T)$ is approximated by
$(T_0/T)^{3/2}$. The speed of light, Planck and Boltzmann constants are $c$, $h$, and $k_B$,
respectively.

The line shape function, $f_\ell$, is given by the Voigt profile with a correction to the Lorentz
far wing, $\chi(\nu-\nu_0)$.
\begin{equation}
f_\ell(\nu-\nu_0; P,T) =  V(\nu_0;\alpha_D,\alpha_L) \; \chi(\nu-\nu_0)
\end{equation}
where
\begin{equation}
\int_{-\infty}^{\infty} f_\ell(\nu) \; d\nu = 1
\end{equation}
Various prescriptions for $\chi$ are in the literature and we use the sub-Lorenztian line shape of
\citet{Campargue2012} that is determined from laboratory data. The following notation is used to
describe the Voigt profile when determining line shape:
\begin{equation}\label{e:Voigt}
{\mathcal V}(x;\sigma,\gamma) = \int_{-\infty}^{\infty} G(x';\sigma)L(x-x';\gamma)dx'
\end{equation}
where
\begin{equation}
G(x;\sigma) \equiv \frac{e^{-x^2/(2\sigma^2)}}{\sigma\sqrt{2\pi}} \quad {\rm and} 
          \quad L(x;\gamma) \equiv \frac{\gamma}{\pi(x^2+\gamma^2)}.
\end{equation}
The integral in Equation~\ref{e:Voigt} can be evaluated with the real part of the complex error
function. The Voigt line shape in terms of physical parameters is given by:
\begin{equation}
V(\nu_0;\alpha_D,\alpha_L) = {\mathcal V}(\sigma,\gamma) \sqrt{\frac{\ln(2)}{\pi}} \alpha_D
\end{equation}
where the dimensionless parameters
\begin{equation}
\sigma = \frac{\nu-\nu_0}{\alpha_D} \sqrt{\ln(2)} \quad{\rm and}\quad \gamma = 
         \frac{\alpha_L}{\alpha_D}\sqrt{\ln(2)}
\end{equation}
are given in terms of the following physical parameters
\begin{equation}
\alpha_D = \frac{\nu_0}{c}\sqrt{\frac{2kT}{m}} \quad{\rm and} \quad \alpha_L = 
			\gamma_{air}\frac{P}{P_0}\left(\frac{T_0}{T}\right)^n
\end{equation}
respectively, using the transition frequency, $\nu_0$, the molecular mass, $m$, temperature, $T$,
together with the reference line broadening half-width, $\gamma_{air}$ that is determined at a
standard pressure $P_0$ and temperature $T_0$, and varies with some temperature dependent exponent,
$n$.

\subsection{Practical Implementation}

The code described for setting up the atmospheric opacity structure and solving the radiative
transfer is original to this work and implemented as a Python package and that is publicly
available\footnote{https://github.com/adamkovics/atmosphere}, including a Python implementation of
CDISORT\footnote{http://www.libradtran.org} (PyDISORT). The Python package management tools should
facilitate installing and compiling the code. Reference data files can be downloaded using methods
within the \texttt{atmosphere} package.

\section{Results}\label{s:Results}

The reference atmospheric model described in Section~\ref{s:RT} is based on the DISR, HASI, and GCMS
measurements. It is most applicable to the tropical regions of Titan during the epoch of the {\it
Huygens} probe descent in 2005. The distribution of aerosol in the atmosphere, which is critical for
the calculation of near-IR spectra, is known to vary on seasonal timescales
\citep[e.g.,][]{Lorenz2004}, however, there is no predictive model for the global vertical structure
of aerosol during the epoch of the observations studied here, so we fit the SINFONI observations to
retrieve the aerosol structure \citep[e.g.,][]{Adamkovics2006}.

Synthetic spectra calculated using models of the aerosol scattering and structure made using DISR
\citep{Tomasko2008c} can overestimate the observed intensities by a factor of up to $\sim$2.
Reducing the aerosol single scattering albedo, $\omega_H$, reducing the aerosol opacity, or changing
the scattering phase function can each reduce intensities in synthetic spectra. \citet{deBergh2012}
and \citet{Griffith2012b} fit VIMS observations in the H-band window by decreasing $\omega_H$ to
0.94 and removing the back-scattering portion of the phase function below 80\,km. That is, they use
the high altitude DISR phase function throughout the atmosphere. The overestimated intensities can
also occur at longer wavelengths (e.g., 2\um) where there is no back-scattering peak in the DISR
aerosol model. The scattering albedo is unconstrained beyond the wavelength range of DISR
observations, so $\omega_H$ can be decreased to reconcile synthetic spectra with observations 
\citep[e.g.,][]{Griffith2012b,Hirtzig2013}. There is significant degeneracy among the aerosol
properties when interpreting spectra, and a critical evaluation of the aerosol scattering is beyond
the scope of this work.

Here we assume that the aerosol single scattering albedo and phase functions, which are related to
the composition and morphology of the particles, are the same as measured by DISR on the {\it
Huygens} probe. We set $\omega_H = 0.96$ and do not vary the phase function, and instead consider
changes in the aerosol optical depth.

We use the vertical profile measured by DISR \citep{Tomasko2008c} as a guide and do not treat the
aerosol opacity in each model layer as a free parameter. That is, we do not consider arbitrary
aerosol vertical structures.  Instead, we consider two cases for variation in the aerosol vertical
structure: (1) We assume that the tropospheric aerosol is uniform below  10\,km and that the aerosol
opacity above this altitude can be scaled linearly from the DISR measurements by a factor, $f_H$.
This approximation treats that spatial variation in haze with a single parameter. (2) In order to
explore the decoupling between tropospheric and stratospheric aerosol, we consider two independent
scale factors for the hazes above and below 65\,km. Neither of these models is motivated by
microphysics, nor do they  take into account empirical evidence for the variations in the aerosol
scattering. However, they are sufficient for interpreting the observations.

Synthetic spectra are compared to observations using a standard quality metric,
\begin{equation}
\hat\chi^2 = \frac{1}{N}\sum_i \left(\frac{r_i}{\sigma_{\rm obs}}\right)^2,
\end{equation}
where the residual between the observation and model, $r_i$, for each data point, $i$, is normalized
by the observational uncertainty, $\sigma_{\rm obs}$, and summed over a total of $N$ data points. We
conservatively estimate that $\sigma_{\rm obs}=0.005$ for the SINFONI observations. A $\hat\chi^2
\lesssim 1$ generally suggests an acceptable model.

Synthetic spectra are generated for the various aerosol models discussed above and compared with
SINFONI observations by calculating $\hat\chi^2$, listed in Table~\ref{t:chi}. Assuming that the
DISR aerosol model for the aerosol scattering and vertical structure applies at all latitudes is
inconsistent with the SINFONI spectra, with $\hat\chi^2\sim5$. Removing the backscattering component
in the aerosol scattering phase function, $P(\theta)$, leads to a better fit, but with
$\hat\chi^2\gtrsim3$. Changing both $P(\theta)$ and $\omega_H$ according to \citet{deBergh2012} is a
further improvement, but still not statistically consistent with the observations, since $\hat\chi^2
\sim 3$. Although synthetic spectra are within $\sim$10\% of the observations near the center of the
disk, the intensities are over-estimated near to the limb. Using the DISR model for scattering and
vertical structure, while scaling the entire column of aerosol by a factor $f_H=0.5$, and assuming
this structure applies at all latitudes is a significant improvement with $\hat\chi^2\approx1$. The
decrease in aerosol column is degenerate with the scattering properties of the aerosol, such that
including a spurious back-scattering peak at low altitudes would cause an over-estimate of the
reduction in aerosol opacity.

\begin{figure*}[]\begin{center}
\iftwocol
	\includegraphics[width=7.0in]{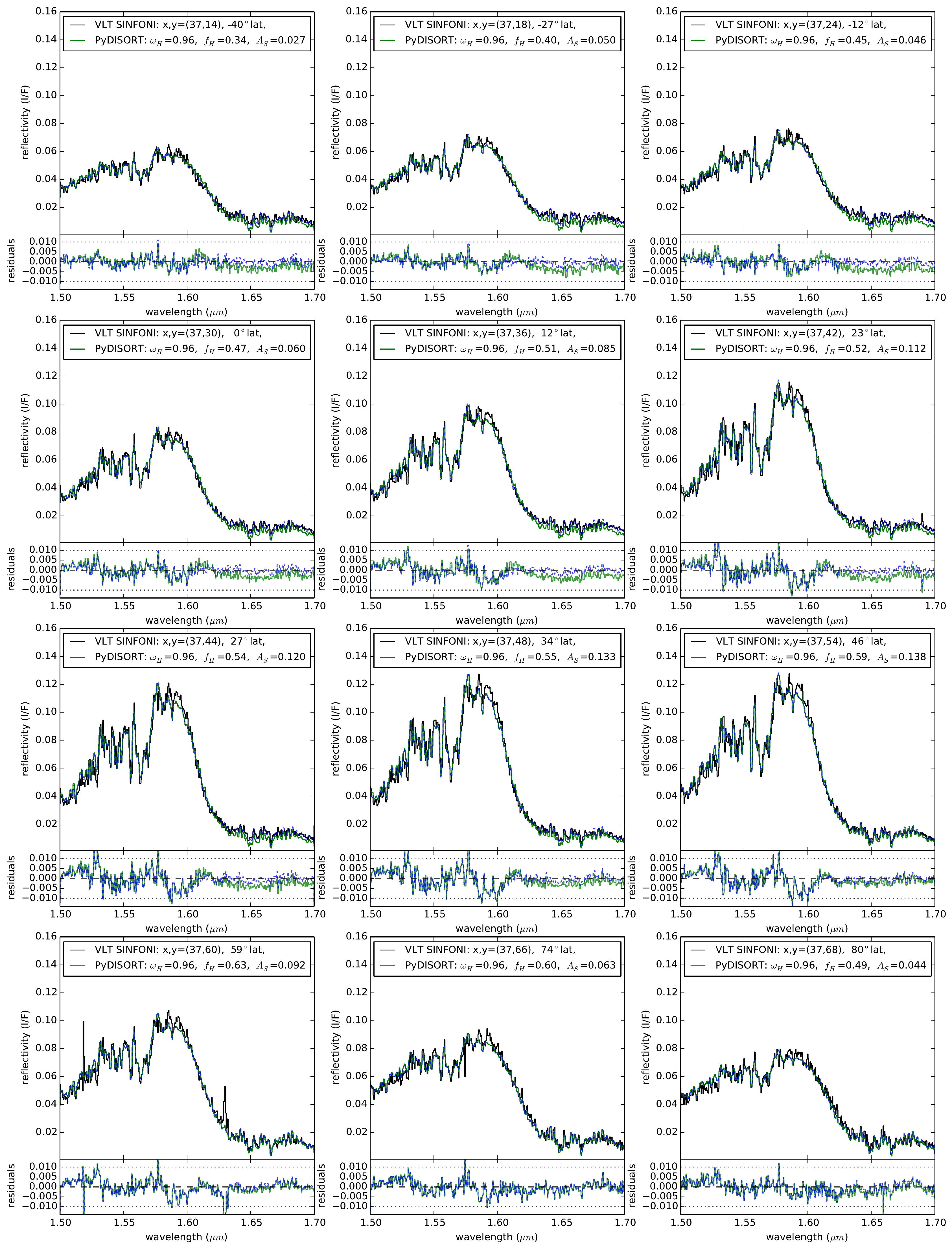}  
\else
	\includegraphics[width=5.0in]{f3}  
\fi
\caption{\label{f:VLT_spectra}VLT SINFONI spectra (black) with best fit 2-parameter models 
(green) at 12 locations in the SINFONI FOV corresponding to the NIRSPAO slit. The free parameters in
the model are the surface albedo, $A_S$, and a scale factor, $f_H$, for the atmospheric haze, which
is known to vary spatially and temporally relative to the haze structure measured by the DISR. A
3-parameter model fit (blue), where scale factors for hazes above and below 65\,km are two separate
free parameters, produces a comparable fit to the observations from 1.50 -- 1.62\um, while fitting
the observations better from 1.63 -- 1.70\um.}
\end{center} \end{figure*}

The hazes on Titan are known to not be uniform, so we fit the SINFONI spectra with two variable haze
models. We use a Levenberg-Marquardt (LM) optimization is to determine the surface albedo, $A_S$,
and either one or two haze scale factors, $f_H$, for each of the SINFONI spectra, which we refer to
as ``2-parameter" and ``3-parameter" models, respectively. The free parameters are well-constrained
with reasonable initial estimates, justifying the LM optimization., SINFONI spectra along the region
corresponding to the NIRSPAO slit are considered, and 12 of these spectra are plotted as examples in
Figure~\ref{f:VLT_spectra}, illustrating that these parameters are sufficient for interpreting the
observations. With $\hat\chi^2 \sim 0.7$ and 0.6 for the 2- and 3-parameter models, respectively,
the variable haze models are significantly better at reproducing the observations than by assuming
the same aerosol vertical profile at all latitudes. The measurements cover a range of viewing
geometries from the center of the disk to the limb, with the limb being most sensitive to the
scattering phase function. The 3-parameter model produces a better fit to the 1.65--1.70\um\ region
(sensitive to the stratosphere) at latitudes south of 20\odeg N, where a comparatively large
decrease in tropospheric aerosol is compensated by a smaller decrease in stratospheric aerosol.

\begin{figure}[t]\begin{center}
\iftwocol
	\includegraphics[width=3.5in]{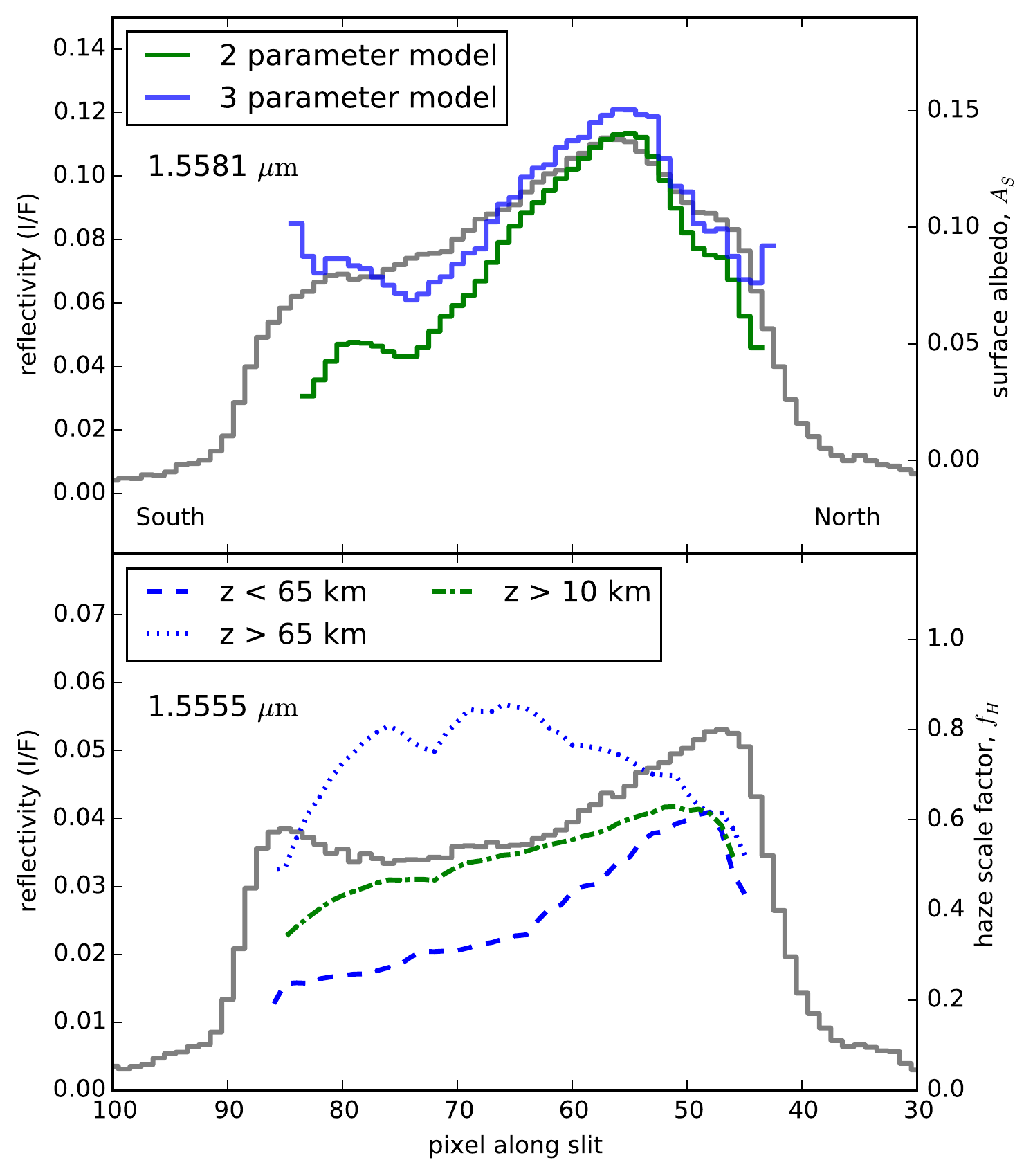}  
\else
	\includegraphics[width=4.5in]{f4}  
\fi
\caption{\label{f:VLT_par}Parameters from LM optimization of model spectra fit to 
SINFONI observations along the slit (a subset are shown in Figure~\ref{f:VLT_spectra}), together
with the NIRSPAO reflectivity at individual wavelengths (black) that are sensitive to these
parameters. The best fit surface albedos (top), where $A_S$ is always one of the free parameters in
the 2 and 3 parameters models (solid green and blue, respectively). And the haze scale factors,
$f_H$, (bottom panel) that are used for either the entire vertical profile above 10\,km (green
dot-dash curve) or as two separate free parameters for the hazes above (blue dotted) and below
65\,km (blue dashed).  The 3-parameter model has a larger gradient in lower atmospheric haze than
the 2-parameter model, and the stratospheric hazes decrease toward both poles. The large haze
gradient in the 3-parameter model is balanced by a small gradient in the surface albedo. The model
parameters have been interpolated onto the NIRSPAO grid and are used as input parameters for fitting
the NIRSPAO spectra.}
\end{center} \end{figure}

The optimized parameters for $A_S$ and $f_H$ from the SINFONI observations are plotted in
Figure~\ref{f:VLT_par}. These values are interpolated onto the NIRSPAO pixel scale and constrain the
NIRSPAO retrievals. The hemispheric asymmetry in $A_S$ that is seen in the SINFONI observations and
the reprojected VIMS map, Figure~\ref{f:view}, is also observed in the NIRSPAO profile at 1.5581\um.
While it is useful to have a benchmark measurement of $A_S$, there are roughly 16\,hrs between the
NIRSPAO and SINFONI observations, over which Titan rotates $\sim$16\odeg, (Table~\ref{t:obslog}).
Due to the timespan between observations, we allow for an offset of the surface albedo, $\delta
A_S$, to be a free parameter in fit of the NIRSPAO observations. The differences in surface albedo
at the times of the two observations can be seen by comparing the red and black dashed lines in
Figure~\ref{f:fluxcal}.

\begin{table}[]
\footnotesize
\caption{Comparison of aerosol models used for SINFONI observations}
\label{t:chi}
\begin{tabular}{lcl} 
\hline\hline
 Model ID & $\hat\chi^2$ & Note \\ \hline
DISR                  & 4.72 & \citet{Tomasko2008c} at all latitudes.  \\ 
$P(\theta)$           & 3.61 & No back-scattering in phase function. \\
$P(\theta), \,\omega_H=0.94$ & 2.86 & No back-scattering, lower scattering albedo. \\
$f_H=0.5$             & 1.02 & Scaled DISR aerosol at all latitudes.  \\
LM 2-parameter        & 0.68 & Variable haze above 10\,km. \\
LM 3-parameter        & 0.57 & Variable haze to surface, and above 65\,km. \\ \hline
\end{tabular}
\end{table}

The interpolated values of $f_H$ are used to generate the atmospheric structure for fitting the
NIRSPAO observations. Once again we use an LM optimization, in this case for three free parameters
for spectra at each pixel along the slit: (1) an adjustment to the input surface albedo, $\delta
A_S$, (2) a high altitude haze factor, $f'_H$, which scales the opacity in the uppermost layer of
the model and accounts for the fact that the 2-parameter model doesn't necessarily fit spectral
regions sensitive to the stratosphere, such as the low reflectance region in the NIRSPAO
observations, and  (3) a scale factor for the tropospheric methane (both
\methane\ and CH$_3$D) abundances that were measured by the GCMS on {\it Huygens}, $f_{\rm CH4}$.

\begin{figure*}[]\begin{center}
\iftwocol
	\includegraphics[width=6.75in]{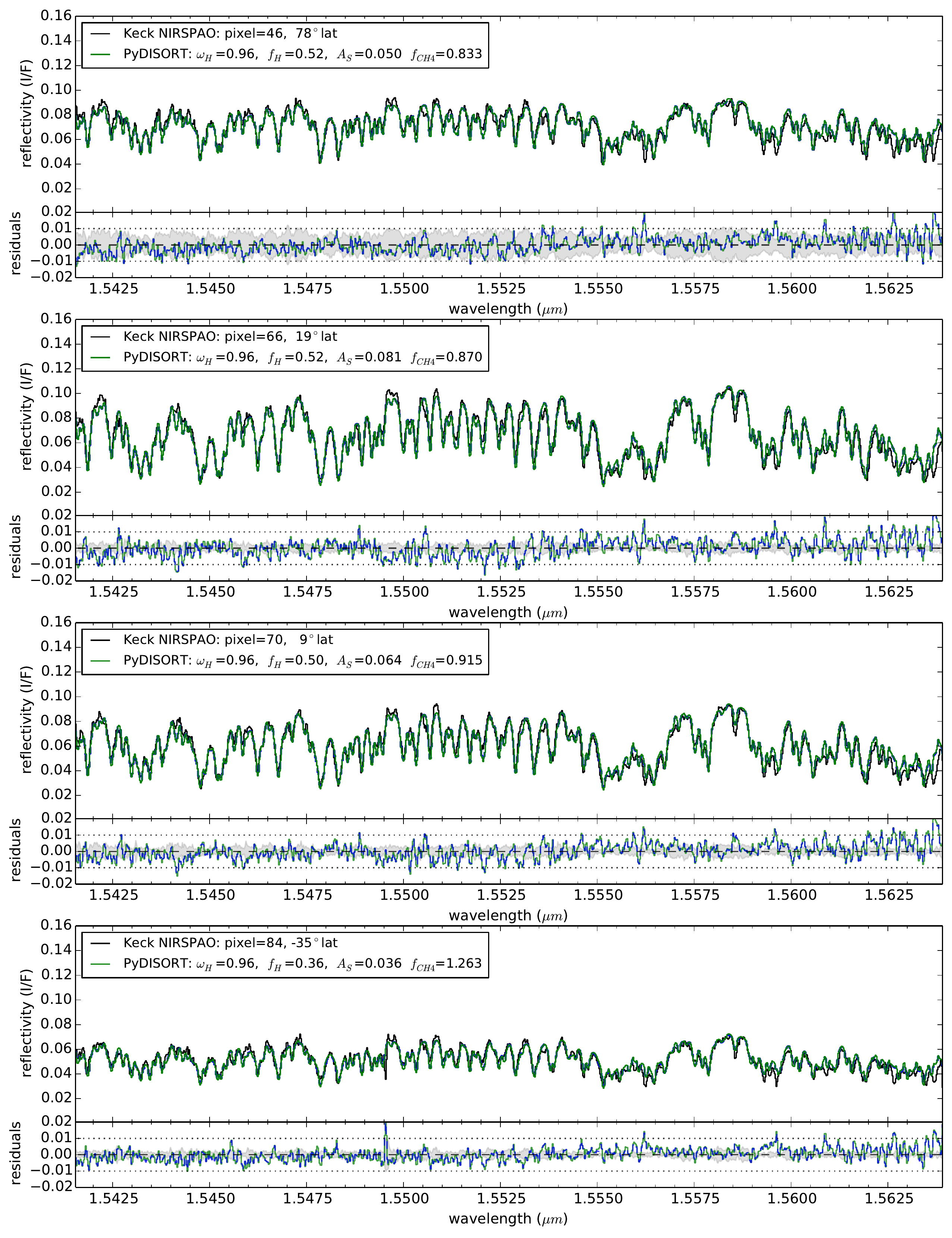}  
\else
	\includegraphics[width=4.5in]{f5}  
\fi
\caption{\label{f:NSPEC_fit}Keck NIRSPAO spectra from four individual pixels along the slit (black) 
with models from a Levenberg-Marquardt optimization using inputs from either 2-parameter (green) or
3-parameter (blue) fitting of SINFONI spectra. The 2-parameter models assume uniform troposphere
haze and only vary the aerosol above 10\,km, whereas the 3-parameter models are optimized to the
observations with a significant gradient in tropospheric aerosol (Figure~\ref{f:VLT_par}). The free
parameters when fitting the NIRSPAO spectra are $f_{\rm CH4}$, $\delta A_S$, and $f_H'$ (see text
for details). These four spectra are a subset of the 42 pixels that sample the disk of Titan and are
used for methane abundance retrievals.}

\end{center}
 \end{figure*}

The best fit spectra from 4 out of 42 pixels on the disk are presented in Figure~\ref{f:NSPEC_fit},
demonstrating the S/N obtained in the observations for one pixel along the slit and the agreement
between models and observations. Both models fit the observations equally well, with residuals that
are generally smaller than the estimated observational uncertainty. The estimate of the
pixel-to-pixel noise in the observations in determined by taking the standard deviation of 5 pixels
centered  on the spectrum of interest. This is an overestimate of noise since the 5 pixels probe
different spatial locations on the disk. This estimate also ignores systematic uncertainties beyond
pixel-to-pixel noise variations. The 1$\sigma$ uncertainties are illustrated as the shaded gray
regions in the residuals panel for each spectrum in Figure~\ref{f:NSPEC_fit}. Parameters for each
model are in the legend.

The LM optimization is performed for all spectra from 42\odeg S to 80\odeg N using 
both haze models, assuming either spatially uniform haze below 10\,km or variable 
haze in this altitude region. Uncertainties are estimated using the roots of the diagonal 
elements of the 3$\times$3 covariance matrix of the LM optimization. A plot of the latitudinal 
variation of $f_{\rm CH4}$ is presented in Figure~\ref{f:methane_var} for both model
assumptions. In the case of uniform haze below 10\,km, there is a significant increase
in tropospheric methane toward the Southern Hemisphere and a slight depletion in the 
Northern temperate regions, to $\sim$90\% of the value measured by the GCMS on {\it Huygens}. If 
the tropospheric aerosol is assumed to be variable, then the methane abundance is essentially
uniform. The degeneracy between the tropospheric haze and methane is due to 
the broadening and blending of methane lines at pressures above a bar. Increasing methane near 
the surface leads to smaller reflectivity at surface-probing wavelengths in a manner 
similar to decreasing tropospheric aerosol. Independently constraining the spatial variation 
in tropospheric aerosol can break the degeneracy between these two models.

\begin{figure*}[]\begin{center}
\iftwocol
	\includegraphics[width=7.0in]{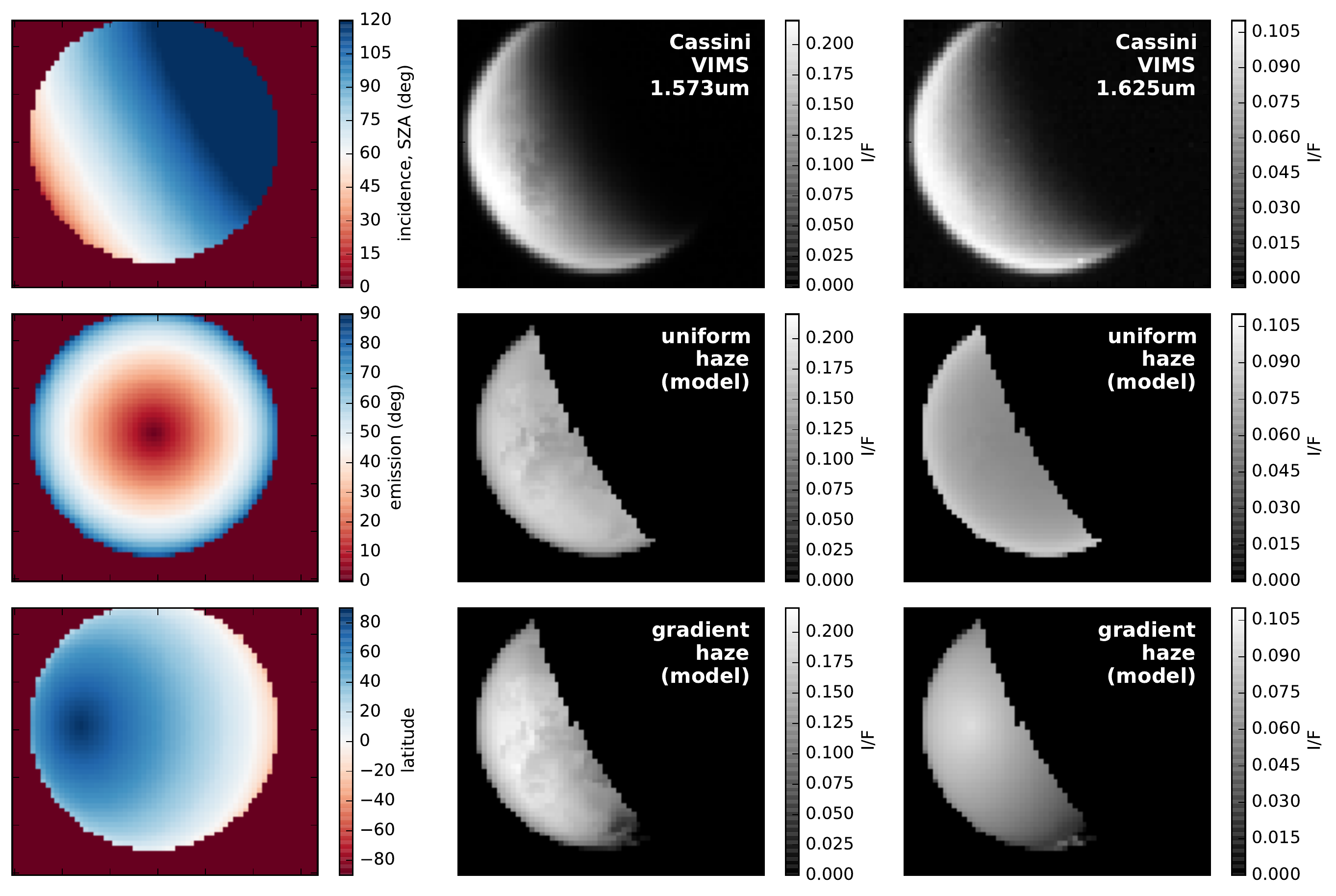}  
	\includegraphics[width=7.0in]{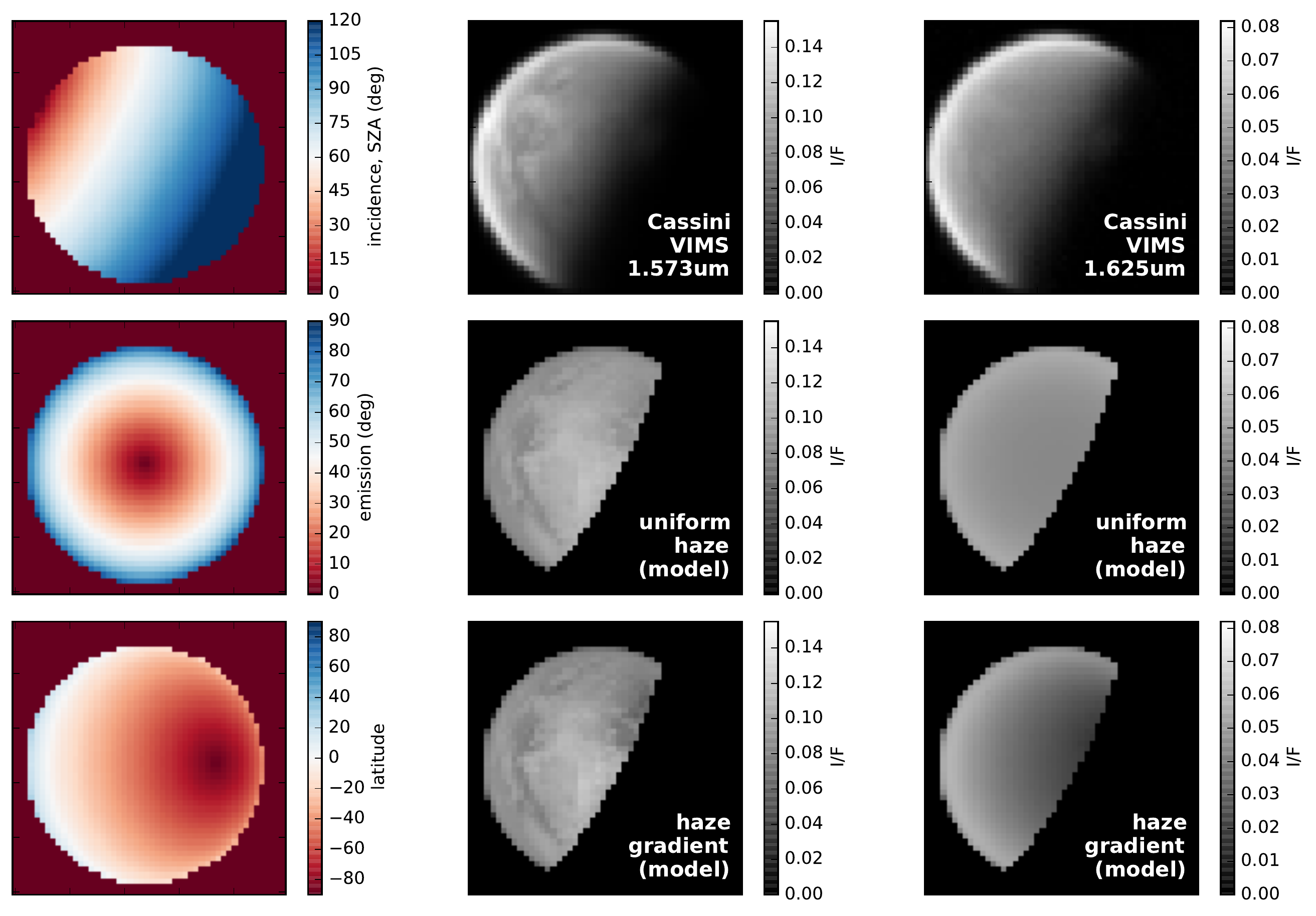}  
\else
	\includegraphics[width=5.25in]{f7} 
	\includegraphics[width=5.25in]{f8}  
\fi
\caption{\label{f:VIMS} {\it Cassini} VIMS images and models (grayscale) of the surface 
and near-surface hazes in the northern hemisphere during the T103 flyby on 2014~Jul~20 (egress).
Viewing geometry is illustrated in the color panels, left column. The synthetic images are
calculated with aerosol models that assume either a uniform haze at all latitudes, or a models with
a haze gradient that increases $f_H$ with latitude toward the North.  }
\end{center} \end{figure*}

Three days after the ground-based observations there was a {\it Cassini} flyby (T103) that passed
over the Southern Hemisphere on ingress and over the Northern Hemisphere on egress, with VIMS
recording global views of the polar regions of Titan. Figures~\ref{f:VIMS} show VIMS images from the
1.5\um\ window in channels that are sensitive to the surface (1.573\um) and a neighboring channel
(1.625\um) where the contribution from surface is small. Differences in reflected intensity due to
surface albedo variation can dominate those due to spatial variation in haze, so we calculate
synthetic images for both channels, considering either uniform haze, or a haze with a gradient of
increasing opacity toward the North.

The synthetic 1.573\um\ images in Figure~\ref{f:VIMS} have decreasing surface albedo contrast toward
the limb and the terminator, due to the increasing path length through the atmosphere associated
with large incidence or emission angles. The greatest albedo contrast is near the dark, polar lakes
in the Northern Hemisphere and near mid-latitudes in the Southern Hemisphere. Assuming a haze
gradient leads to a brighter region near the pole, and slightly greater surface albedo contrast near
the South Pole. In general, there are only minor differences in the synthetic images of the surface
when comparing the two haze models.

The synthetic 1.625\um\ images, however, illustrate an increased sensitivity to the aerosol model. A
uniform haze leads to limb brightening that is consistent with observations both in the North and
the South polar regions, whereas the gradient in the haze leads to brighter Northern, and a darker
Southern, polar regions than are observed by VIMS. A uniform distribution of aerosol near the
surface means that the NIRSPAO spectra and are more consistent with a meridional variation in
methane, and not a variation in aerosol.

\section{Discussion}\label{s:Discussion}

We have used complementary ground-based observations from NIRSPAO at Keck and SINFONI at the VLT to
measure the tropospheric methane distribution on Titan. The spatially-resolved SINFONI observations
at moderate resolution, across the entire H-band, are used for the flux calibration and constraining
the aerosol haze distribution, while the high resolution NIRSPAO observations are sensitive to the
tropospheric methane. We used HITRAN 2012 line lists to generate $k$ distributions for \methane\ and
CH$_3$D at the resolutions and plate scales that correspond to each of the observations.
Levenberg-Marquardt optimization was used to fit the observations, and retrieve the tropospheric
methane abundance, assuming either variable or uniform tropospheric haze. {\it Cassini} VIMS images
suggest that the haze is uniform toward both poles and that there is meridional variation in
methane.

\begin{figure}[]\begin{center}
\iftwocol
	\includegraphics[width=3.5in]{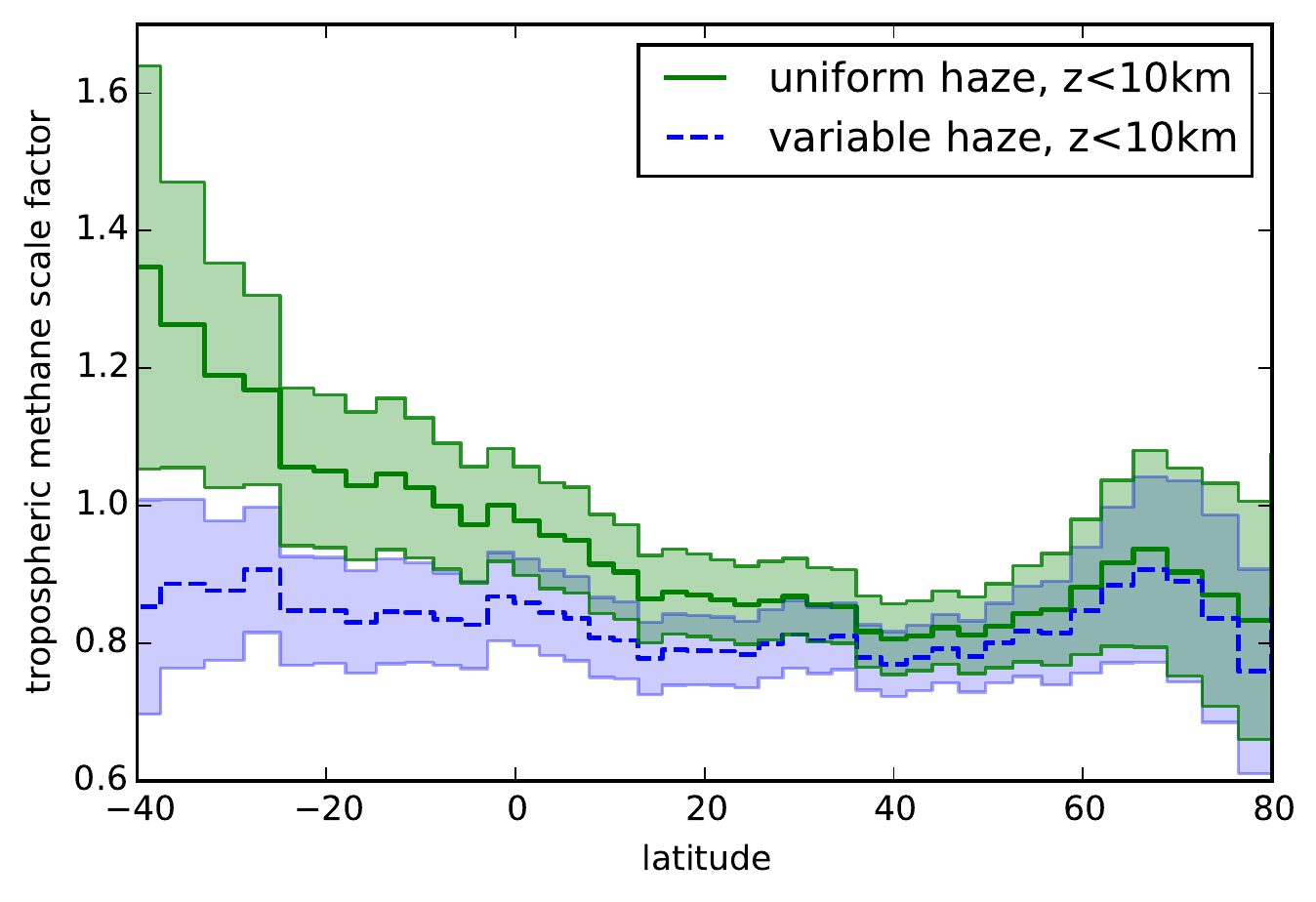}  
\else
	\includegraphics[width=5.25in]{f6}  
\fi
\caption{\label{f:methane_var}
Meridional variation in the tropospheric methane abundance relative to the mole fraction measured by
the GCMS on the {\it Huygens} probe retrieved from fits to NIRSPAO spectra. Examples spectra and
best-fit models are illustrated in Figure~\ref{f:NSPEC_fit}.}
\end{center} \end{figure}

The distribution of methane in Figure~\ref{f:methane_var} reveals two interesting features. First,
it is nearly uniform in the northern hemisphere from ~15$^{\circ}$\,N to the pole, and marginally
lower ($\sim$10\%) than the Huygens GCMS measurement.  Second, concentrations rise monotonically to
the south of 15$^{\circ}$\,N and reach a peak value of 1.2-1.6 times the Huygens GCMS measurement at
40$^{\circ}$\,S.  This suggests that there are at least two, distinct source regions of methane
vapor, and that the atmosphere is mixing the air masses from these source regions in latitudes south
of 15$^{\circ}$\,N.  If we assume the source regions produce saturated air masses at the local
temperature, we can estimate the difference in their temperatures.  For a given temperature
difference, $\Delta T$, the Clausius-Clapeyron relation predicts a fractional change in saturation
vapor pressure, $\Delta e_s/e_s \approx L_v\Delta T/(R_v T^2)$, were $L_v$ is the latent heat of
vaporization and $R_v$ is the methane gas constant.  Assuming the northern hemisphere source region
has a temperature of $\sim$91\,K, a 40\% change in the saturation vapor pressure requires the source
regions have a temperature difference of $\sim$3 K, which is approximately the largest observed
surface brightness temperature difference \citep{Jennings2011}.

Evaporation from Ontario Lacus \citep{Turtle2011b,Hayes2009} as the sole explanation for methane
enhancement (during southern summer) was rejected by \citet{Tokano2014} based on the small spatial
coverage of Ontario Lacus (0.04\% of the Southern hemisphere) and the assumption of $\sim$4\,m
change in lake depth. This suggests that lake evaporation at the poles over the timescale of the
{\it Cassini} mission is an unlikely explanation for the increase in methane from the equator toward
southern mid-latitudes at the onset of southern winter. However, evaporation from moist ground or a
number of small, methane-dominated lakes is still possible. Indeed, evaporation rates from moist
surfaces can exceed the rates from standing liquids due to the larger surface area available and the
possibility for increased turbulence above a rough surface. Our results suggest that evaporation
from the surface, in a region poleward of a particular latitude, is increasing the relative humidity
of the atmosphere. The observed gradient in methane abundance (toward the winter pole) may then be
an indication of a relatively moist air parcel from the pole equilibrating to the drier conditions
at the equator during meridional transport.

In November 2000, \citet{Anderson2008} used spectral images from 0.6 -- 1.0\um, obtained with the
{\it Hubble Space Telescope} (HST), to measure a latitudinal gradient from the south pole toward the
northern (winter) mid-latitudes. They report that the tropospheric methane column roughly doubled
from $\sim$70\odeg S to $\sim$10\odeg N, and were presumably sensitive to to the near surface
humidity. Since the HST observations were nearly a half Titan-year earlier, the seasons should be
analogous to our ground-based observations, but mirrored North to South. One difference between the
measurements is that \citet{Anderson2008} report a significant drop in the methane column in their
northern-most datapoint at $\sim$30\odeg N, whereas our measurements increase through $\sim$40\odeg
S. Another consideration is that \citet{Anderson2008} report a much weaker, or absent, latitudinal
variation in methane 7 days earlier at a different central meridian longitude (CML), which they
speculate is an indication of either a surface or sub-surface source of methane that is spatially
variable.

An increase in methane toward the winter polar is at odds with the predictions of circulation models
that have been used to interpret {\it Cassini} radio occultation data. The measurements obtained
over the period from southern summer to southern equinox (2006-2009) indicate that tropospheric
methane should increase from the equator  toward the Summer pole \citep{Tokano2014}. The latitudinal
gradient in sea-level pressures that are retrieved by \citet{Tokano2014}, assuming that methane
increases toward the Summer pole, are consistent with the location of the seasonal convergence zone
and the observed regions of precipitation as tracked by cloud formation \citep{Mitchell2012}.
Further work is required to determine conclusively if these observations test particular assumptions
of circulation models, e.g., the distribution of methane at the surface, or if some other mechanism
needs to be invoked to interpret the measurements. For example, episodic releases of methane, from
some form of cryovolcanism, have been suggested as mechanism for supplying the moisture to southern
mid-latitude clouds \citep{Roe2005b}. However, the rate of cloud occurrence in the South has dropped
with the changing seasons and the locations of clouds are generally thought to be controlled by
circulation, rather than topography \citep{Roe2012, Mitchell2012}. Nonetheless, the properties of
the surface regolith remain a mystery and the potential for the episodic release of methane near
polar latitudes is unconstrained.

Repeated ground-based observations in different epochs, and at different CML can test for
contemporary variability is near-surface sources of methane. If mixing of saturated polar air during
transport controls the gradient in tropospheric methane, we can predict that the gradient in methane
should change seasonally with the changing circulation. Changes in the gradient on more rapid
time-scales, or at different CML can confirm either episodic releases of methane or spatial
variation in the source region. Simultaneous observations with IFU and slit spectrometers, and
quantitative constraints on the surface albedo from VIMS maps could further constrain our models.
Degeneracies in the parameterized properties of the haze could also be further constrained by
aerosol microphysical models that predict how the vertical structure of aerosol opacity and
scattering phase functions change with time. Continuing work on methane line assignments, line
shapes, and possible variations in upper stratospheric methane
\citep[e.g.,][]{Lellouch2014} can be used to make quantitative improvements 
to both the relative and absolute uncertainties in the retrieval.

We have presented measurements of the meridional variation in tropospheric methane on Titan. These
results suggest that localized regions of evaporation occur at Southern polar latitudes in the
winter, likely from a moist regolith. The simultaneous analysis of Keck NIRSPAO, VLT SINFONI, and
{\it Cassini} VIMS observations illustrate the challenges in performing an accurate retrieval. We
have mentioned a few improvements for constraining assumptions in our radiative transfer models and
have suggested additional observations of this type at future epochs. Accurately measuring the
spatial variation in the tropospheric methane will constrain sources of methane at the surface and
inform our understanding of the hydrological cycle on Titan.

\section*{Acknowledgements}
This work was supported by NASA PAAST grants NNX14AG82G and NNX12AM81G. M\'A was supported in part
by NSF AST-1008788. PMR was supported by Fondecyt grant \#1120299. We wish to acknowledge
Jonathan~I. Lunine and Elizabeth~P.~Turtle, who are members of the VLT SINFONI
cloud observing campaign that provided the SINFONI observations presented here. Some of the data
presented were obtained at the W.~M.~Keck Observatory, which is operated as a scientific partnership
among the California Institute of Technology, the University of California and the National
Aeronautics and Space Administration. The Observatory was made possible by the generous financial
support of the W.~M.~Keck Foundation. The authors wish to recognize the significant cultural role
that the summit of Mauna Kea has always had within the indigenous Hawaiian community. We are
fortunate to have the opportunity to conduct observations from this mountain.

\section*{References}


\end{document}